

\input phyzzx.tex
\PHYSREV

\def\BH{black hole}

\def\os{\beta=\beta_{h}}

\def\pano{\par \noindent}
\def\sch{Schwarzschild}

\def\text{Tr_{+}}

\title{Entropy and topology for manifolds with boundaries}
\author{Stefano Liberati\foot{E-mail:liberati@roma1.infn.it}
and
Giuseppe Pollifrone\foot{E-mail:pollifrone@roma1.infn.it }}
\address{Dipartimento di Fisica, Universit\`a di Roma ``La
Sapienza"
 \pano and INFN, Sezione di Roma, Piazzale Aldo Moro 2,
00185 Roma, Italy}

\abstract{In this work
a deep relation between topology and thermodynamical
features
of manifolds with boundaries is shown.
The expression for the Euler characteristic, through the Gauss-
Bonnet
integral,
and the one for the entropy of gravitational instantons are
proposed in a form
which makes the relation between them self-evident.
A generalization of Bekenstein-Hawking formula, in which
entropy
and
Euler characteristic are related in the form $S=\chi A/8$, is
obtained.
This formula reproduces the correct result for extreme \BH,
where the
Bekenstein-Hawking one fails ($S=0$ but $A \neq 0$).
In such a way it recovers a unified picture for the \BH\
entropy law.
Moreover, it is proved that such a relation can be generalized to
a wide class
of manifolds with boundaries which are described by
spherically symmetric
metrics (e.g. \sch\ , Reissner-Nordstr\"{o}m, static de Sitter)}.

\chapter{Introduction}

Recent works by Hawking, Horowitz and Ross
\Ref\HHi{S.W.~Hawking,
G.T.~Horowitz S.F.~Ross -
{\em Entropy, Area and black hole pairs} - Preprint gr-qc
9409013.},
\Ref\HHii{S.W.~Hawking, G.T.~Horowitz - {\em The
gravitational
hamiltonian,
action, entropy and surface terms} - Preprint gr-qc 9501014.}
have demonstrated that usual Bekenstein-Hawking law for
\BH\ entropy
fails in
the case of extreme \BH\ . For these kinds of object we have a
null entropy in
spite of a non null area of the event horizon. These authors
observed that this
change in extreme case in respect of non-extreme one is mainly
due to the
different nature of event horizon in the former.
One infact finds that in these cases, the presence of the event
horizon is not
associated with a non-trivial topology of spacetime.
Euler characteristic is infact zero (not two) for this kind of
\BH\ .
This radical difference in extreme \BH\ physics seems a strong
hints
towards
a point of view that particular case of \BH\ solutions
(for example
extreme Reissner-Nordstrom \BH\ is ``just" the case
$Q^{2}=M^{2}$ of the
general solution) is to be considered a rather different object from
the non-extreme one.
The extreme \BH s are interesting in their own right because of their very
special topological structure and because they appear to be the
limit in which
one loses intrinsic thermodynamic of black hole. They are supposed
to have a
physical
interpretation as elementary particles of fundamental theories
of gravity
\Ref\Sen{A.~Sen - Preprint hep-th/9504147} so their
understanding appears
to be
an important point in developement of a theory of quantum gravity.
Nevertheless we think that it is possible to understand extreme case as a
particular case
of \BH\ without requiring a limitation of \BH\
thermodynamics laws.
The guiding idea (originally proposed by Gibbons and Hawking
\Ref\GH{G.W.~Gibbons, S.W.~Hawking\journal
Phys.Rev.&D15, 10 (1977)
2752.}) of this work is
that thermodynamical features of spacetimes like the \sch\
one are
explainable as an effect due to their non-trivial topological
structure and in
particular to the nature of their boundaries.
We will see in particular that Euler characteristic and entropy
have the same
dependence on the boundaries of the manifold and
we will relate
them in a general formula. This relation (although
demonstrated only for a
certain class of metrics) would be valid for every compact
manifold on which
Gauss-Bonnet theorem can be extended.

\chapter{Euler characteristic and manifold structure}

The Gauss-Bonnet theorem proves that it is possible to obtain
the Euler
characteristic of a 4-dimensional compact riemannian manifold
$M$ without
boundaries by the volume integral of the 4-dimensional metric
curvature:
$$
S_{GB}={{1}\over{32 \pi^{2}}} \int_{M} \epsilon_{abcd}
R^{ab}\wedge R^{cd}
$$
with $R$ bound to the spin-connections $\omega$ of the
manifold by the
relations:
$$
R^{a}_{b}=d\omega^{a}_{b}+\omega^{a}_{c}\wedge
\omega^{c}_{b}
$$
Chern\Ref\CHI{S.~Chern\journal Annals of Math.&45(1944)
4, 747.}
\Ref\CHII{S.~Chern\journal Annals of Math.&46(1945) 4,
674.}
showed that the differential n-form $\Omega$ of Gauss-
Bonnet integral:
$$
\Omega={{(-1)^{p}}\over{2^{2p}\pi^{p}p!}} \epsilon_{a_{1}
\ldots a_{p}} R^{a_{1} a_{2}} \wedge \ldots \wedge
R^{a_{2p+1} a_{2p}}
$$
defined on $M^{n}$ can be defined on a manifold $M^{2n-1}$
which is the
image
of $M^{n}$ through the flux of its unitary vectors field.
Then he was able to express $\Omega$ has the exterior
derivative of a
differential $n-1$-form in $M^{2n-1}$:
$$
\Omega=d\Pi
$$
He also demonstrated that the original $\Omega$ integral on
$M^{n}$ can be
performed on a submanifold $V^{n}$ of $M^{2n-1}$ which has
as boundaries
the
set of singular points of the unitary vector field previously
cited.
By Stoke's theorem we then obtain:
$$
S^{vol}_{GB}=\int_{M^{n}}\Omega=\int_{V^{n}}\Omega=
\int_{\partial V^{n}}\Pi
$$
For manifold with boundaries this formula can be
generalized \Ref\EGH{T.~Eguchi, P.B.~Gilkey, A.J.
Hanson\journal Phys.
Rep.
&66, 6 (1980)}:
$$
S_{GB}=S^{vol}_{GB}+S^{bou}_{GB}=\int_{M^{n}}\Omega-
\int_{\partial M^{n}}\Pi=\int_{\partial V^{n}}\Pi-
\int_{\partial M^{n}}\Pi
\eqn\Sgb
$$
This expression implies that the Euler characteristic of a
manifold $M^{n}$
with boundaries becomes null in the case that its contours
would be the same
as
that of the submanifold $V^{n}$ of $M^{2n-1}$.\pano
We shall now use \Sgb\ for \BH\ manifold. We always work
in Euclidean
manifolds after imaginary time compactification necessary in
order to remove
conical singularities on the horizons.

{ \bf{ $\chi_{euler}$ of no-extreme \BH\ }}
\pano
For no-extreme \BH\ the boundaries of the manifold $V$
are set by the extreme values of the range of radius coordinate
that are
$r=r_{h}$ and $r=r_{0}=\infty$.
The physical manifold $M$ instead has just one boundary at
infinity because,
after removal of conical singularity, the \BH\ horizon
$r=r_{h}$ is not a
border of spacetime.
So:
$$
\Sgb=\int_{r_{0}}\Pi -\int_{r_{h}} \Pi -\int_{r_{0}} \Pi=
-\int_{r_{h}} \Pi
$$
It is possible to use this formula for calculate Euler number and
for example
in Schwarzschild case one correctly
obtains $\chi_{euler}=2$ which is the expected value for
$S^{2}\times R^{2}$
topology.\pano

{ \bf{$ \chi_{euler} $ for extreme \BH\ }}
\pano

For extreme \BH\ the boundaries of the manifold $V$
are the same as that one for the ordinary case $r=r_{h}$ and
$r=r_{0}=\infty$.
On the other hand the physical manifold $M$ has now two
boundaries at
infinity
represented by the usual spatial infinity $r=r_{0}=\infty$ and by
the horizon
$r=r_{h}$.
In fact, in this case the time-affine Killing vector
has a set of fixed
points only at infinity (it becomes null only asymptotically at
infinity, in this
sense one says that the \BH\ horizon for extreme \BH\ is at
infinity).
So:
$$
\Sgb=\int_{r_{0}}\Pi -\int_{r_{h}} \Pi -\int_{r_{0}} \Pi
+\int_{r_{h}} \Pi=0
$$
This shows that for extreme \BH\ the Euler characteristic is
always null.

\chapter{Entropy for manifolds with boundaries}

We will follow the definition of \BH\ entropy adopted by
Kallosh, Ortin,
Peet \Ref\KOP{R.~Kallosh, T.~Ortin, A.~Peet\journal Phys.
Rev.&D47 (1993) 12,
5400.}.\pano

We consider a thermodynamical system with conserved
charges $C_{i}$ and
relative potentials $\mu_{i}$ so we work in grandcanonical
ensemble.
$$
\eqalign {
Z &={\rm{Tr}}\; e^{-(\beta H -\mu_{i} C_{i})} \cr
Z &=e^{-W}\cr
W &=E-TS-\mu_{i} C_{i} \cr }
$$
We obtain:
$$
S=\beta(E-\mu_{i}C_{i})+\ln Z
$$
Gibbons-Hawking demonstrated that at the tree level:
$$
\eqalign{
Z& \sim e^{-I_{E}}\cr
I_{E}&={{1}\over{16 \pi}} \int_{M}(-
R+L_{matter})+{{1}\over{8\pi}}
\int_{\partial M} \left [ K \right ] \cr }
$$
Here $I_{E}$ is the ``on-shell'' Euclidean action.\pano
In calculating $Z$ and so $I_{E}$ it is important to correctly
evaluate the
boundaries of our manifold M.\pano
For no-extreme \BH\ we have just one boundary at infinity
$r_{0}
\rightarrow \infty $ (after the removal of conical singularity, the
metric is regular
on the horizon $r=r_{h}$).\pano
For extreme \BH\ we have a drastic change in boundaries
structure.
Metrics do not present conical singularity so we cannot fix
imaginary time
value. The horizon is at an infinite distance from the external
observer and so
it is as an ``internal" boundary of our spacetime (we can say
that the
coordinate of this internal boundary is $r_{b}$).

In order to determine $S$ we have also to compute
$\beta(E-\mu_{i}C_{i})$.\pano
{}From Gibbons-Hawking [\GH] we know that for two fixed
hypersurfaces at
$\tau=cost$
($\tau=$imaginary time),
$\tau_{1}$ e $\tau_{2}$, one has:
$$
\langle \tau_{1} | \tau_{2} \rangle =e^{-(\tau_{2}-\tau_{1})(E-
\mu_{i} C_{i})}
\approx e^{-I_{E}}
$$
In this case it is necessary to understand that the time-affine
Killing vector
$\partial/\partial \tau$ has two sets of fixed points, one at
infinity the other
on the horizon. So an hypersurface at $\tau=cost$ has two
boundaries in
corresponding to this sets, independently of the position of
horizon (which
can be at infinity for extreme \BH\ ).

So one obtains\foot{Note that, for metrics under our
consideration,
$ V_{bulk}=M_{bulk} $  so the bulk part of the entropy always
cancels also for
metrics which are not Ricci-flat (as de Sitter case). All the
entropy depends
on boundary values of extrinsic curvature.}:
$$
\eqalign{
S&=\beta(E-\mu_{i} C_{i}) +\ln{Z}=\cr
&={I_{E}}^{\infty}_{r_{h}}-{I_{E}}^{\infty}_{r_{boun}}\cr
&={{1}\over{8 \pi}} \left (\int_{\partial V}
[K] - \int_{\partial M} [K ] \right)=\cr
&=\left ( \int_{r_{0}} [K]- \int_{r_{h}} [K]-\int_{r_{0}}
[K]+\int_{r_{b}} [K]
\right )
\cr}
\eqn\entr
$$
The deep similarity between \entr\ and \Sgb is self-evident.

{\bf{Entropy for no-extreme \BH\ }}
\pano
In this case we don't have an internal boundary for $M$ and so
we don't have
$r_{boun}$ in \entr\ and we obtain:
$$
S={{1}\over{8 \pi}} \left [ \int_{\infty} [K]- \int_{r_{h}}[K]
- \int_{\infty} [K] \right]=-\int_{r_{h}}[K]=A/4
$$

{\bf{Entropy for extreme \BH\ }}
\pano
In this case the horizon is at infinity and $M$ has two
boundaries in
$r=\infty$
and $r_{b}=r_{h}$:
$$
S={{1}\over{8 \pi}} \left [ \int_{\infty} [K]- \int_{r_{h}}[K]
- \int_{\infty} [K] +\int_{r_{h}}[K]  \right] =0
$$

Some comments on derivation of \entr\ are in order.
It was in fact derived in a grandcanonical ensemble but for
extreme
\BH\ there is no conical singularity so there is no $\beta$
fixing and consequently there is
no intrinsic thermodynamic of the manifold. We conjecture
that the correct
procedure we have to follows is exactly the inverse. The last
line of \entr\ is
the general expression of entropy for manifold with
boundaries. The lack of
intrinsic thermodynamics is deducible from \entr\ by
consideration of
boundaries
structure. It is not possible to fix $\beta$ because boundary
changes in
extreme
case, not the contrary. Thus \entr\ is generalizable to a large
class of riemannian manifolds with boundaries and the
similarity in boundary
dependence with Gauss-Bonnet integral is a strong hint toward
the evidence of a
link between
entropy and topology for gravitational instantons \foot{As a
seminal work in this
direction we can quote the classical paper by Gibbons and
Hawking about
gravitational instantons \Ref\GHI{G.W.~Gibbons,
S.W.~Hawking\journal Commun. Math. Phys.&66 (1979)
291}.}.

\chapter{Euler characteristic for \sch -like metrics}

We consider metrics of the form:
$$
ds^{2}=-e^{2U(r)}dt^{2}+e^{2U(r)}dr^{2}+R^{2}(r)d^{2}\Omega
\eqn\metr
$$
Spin connections are:
$$
\eqalign{
\omega^{01}&={{1}\over{2}} ( e^{2U})^{\prime} dt\cr
\omega^{21}&=e^{U}R^{\prime}d\theta \cr
\omega^{31}&=e^{U}R^{\prime}\sin \theta d \phi \cr
\omega^{32}&=\cos \theta d \phi \cr}
\eqn\spc
$$
One has that \Ref\GK{G.W.~Gibbons, R.E.~Kallosh-
{\em Topology, entropy and Witten index of dilaton black
holes}
- Preprint hep-th 9407118.} :
$$
S^{vol}_{GB}={{1}\over{32 \pi^{2}}} \int_{M} \epsilon_{abcd}
R^{ab} \wedge R^{cd}={{1}\over{4 \pi^{2}}} \int_{V}
d(\omega^{01} \wedge R^{23})={{1}\over{4 \pi^{2}}}
\int_{\partial V} \omega^{01} \wedge R^{23}
\eqn\gbvo
$$
For the boundary term one finds [\EGH ]:
$$
S^{boun}_{GB}=-{{1}\over{32 \pi^{2}}} \int_{\partial M}
\epsilon_{abcd} (2\theta^{ab} \wedge R^{cd}-{{4}\over{3}}
\theta^{ab} \wedge \theta^{a}_{e} \wedge \theta^{eb})
\eqn\gbbo
$$
{}From \gbbo\ one obtains after some standard algebra [\GK] :
$$
S^{boun}_{GB}=-{{1}\over{4 \pi^{2}}} \int_{\partial M}
\omega^{01} \wedge R^{23}
\eqn\gbbo2
$$
Hence one has the complete result:
$$
S_{GB}=S^{vol}_{GB}+S^{boun}_{GB}={{1}\over{4 \pi^{2}}}
\left ( \int_{\partial V}-\int_{\partial M} \right )
\omega^{01} \wedge R^{23}
$$
For metrics of no-extreme \BH\ one has:
$$
S_{GB}=S^{vol}_{GB}+S^{boun}_{GB}=-{{1}\over{4 \pi^{2}}}
\left (\int_{\partial M} \omega^{01} \wedge R^{23} \right
)_{r=r_{h}}
$$
For the metrics \metr\ :
$$
\eqalign{
R^{23}&=d\omega^{23}+\omega^{21} \wedge \omega^{13}=(1-
e^{2U}(R^{\prime})^{2}) \sin \theta d \theta d \phi \cr
\omega^{01} \wedge R^{23}&={{1}\over{2}} ( e^{2U} )^{\prime}
(1-e^{2U}(R^{\prime})^{2}) \sin \theta d \theta d \phi dt\cr}
$$
As previously said, we shall perform our calculations for
riemannian manifolds
with compactification of imaginary time, $0 \leq\tau
\leq\beta$
(generalization of conical singularity remotion condition for
our class of
metrics).
It is easy to see that this corresponds to the choice:
$$
\beta=4 \pi \left ( (e^{2U})^{\prime}_{r=r_{h}} \right )^{-1}
\eqn\b
$$
Note that condition \b\ gives an infinite range of time (no
period) for the case of extreme \BH\ metrics that gives
$\left ( (e^{2U})^{\prime}_{r=r_{h}} \right )=0$.
For a general manifold $K$ with boundary
$\partial K=(r_{ext},r_{int})$ one has:
$$
\eqalign{
S_{GB}&={{1}\over{4 \pi^{2}}} \int_{\partial K} \omega^{01}
\wedge R^{23}=\cr
&=-{{1}\over{4 \pi^{2}}}\int {{1}\over{2}} (e^{2U})^{\prime}
\sin \theta (1-(e^{U} R^{\prime})^{2} ) d \theta d \phi dt=\cr
&= -{{1}\over{2 \pi}} \left.\beta (e^{2U})^{\prime} (1-(e^{U}
R^{\prime})^{2} ) \right|_{r_{int}}^{r_{ext}}=\cr
&=2 \left ( (e^{2U})^{\prime}_{r=r_{h}} \right )^{-1} \left [
\left.(e^{2U})^{\prime} (1-(e^{U} R^{\prime})^{2} )
\right|_{r_{int}}
-\left.(e^{2U})^{\prime} (1-(e^{U} R^{\prime})^{2} )
\right |_{r_{ext}}\right ] \cr}
\eqn\Sgen
$$
if we consider the manifolds $V$ ed $M$ with boundaries
$\partial
V=(r_{0},r_{h})$ and $\partial M=(r_{0},r_{b})$ we then obtain:
$$
\eqalign{
S_{GB}&=2 \left ( (e^{2U})^{\prime}_{r=r_{h}} \right )^{-1}
\left [ \left( (e^{2U})^{\prime} (1-(e^{U} R^{\prime})^{2}
\right )_{r_{h}} -\left ((e^{2U})^{\prime} (1-(e^{U}
R^{\prime})^{2} ) \right)_{r_{0}}\right ]+
\cr
&-2 \left ( (e^{2U})^{\prime}_{r=r_{h}} \right )^{-1}
\left [ \left( (e^{2U})^{\prime} (1-(e^{U} R^{\prime})^{2}
\right )_{r_{b}} -\left ((e^{2U})^{\prime} (1-(e^{U}
R^{\prime})^{2} ) \right)_{r_{0}} \right]=
\cr
&=2 \left ( (e^{2U})^{\prime}_{r=r_{h}} \right )^{-1}
\left[
\left ((e^{2U})^{\prime} (1-(e^{U} R^{\prime})^{2}
\right)_{r_{h}} - \left( (e^{2U})^{\prime}
(1-(e^{U} R^{\prime})^{2} ) \right)_{r_{b}} \right ]=
\cr
&=2\left [1-(e^{U} R^{\prime})^{2} \right ]_{r_{h}}-
\left ( (e^{2U})^{\prime}_{r=r_{h}} \right )^{-1}
\left [ (e^{2U})^{\prime} (1-(e^{U} R^{\prime})^{2} )
\right]_{r_{b}}
\cr}
\eqn\Sgbg
$$
{}From \Sgbg\ we can see that for no-extreme \BH\ (no $r_{b}$)
one obtains:
$$
S^{BH}_{GB}=2 \left [1-(e^{U} R^{\prime})^{2} \right ]_{r_{h}}
\eqn\GBnBH
$$
For extreme \BH\ ($r_{b}=r_{h}$) one straightforwardly finds:
$$
S^{BH_{extr}} \equiv 0
\eqn\GBeBH
$$

\chapter{Entropy for \sch -like metrics}

We consider again metrics of the form:
$$
ds^{2}=- e^{2U(r)} dt^{2} + e^{- 2U(r)} dr^{2} + R^{2}(r)
d^{2}\Omega
$$
One has:
$$
S=\beta \left(E-\mu_{i}C_{i} \right) +\ln
Z={I_{E}^{\infty}}_{r_{h}}-
{I_{E}^{\infty}}_{r_{boun}}={{1}\over{8 \pi}} \left
(\int_{\partial V}
[K] - \int_{\partial M} [K ] \right)
\eqn\Sgen
$$
$\partial V$ here stays as the border of the manifold $V$  on
which we
consider the surfaces at $\tau=cost$ in order to calculate the
factor
$\beta (E-\mu_{i}C_{i} )$  (these are delimited, for every kind
of \BH\ , by
$r_{h}$ e $r_{0}=\infty$). \pano
$\partial M$ is the boundary of space-time (in this case we
have that the
boundary surfaces are at $r_{h}$  and $\infty$ for extreme
\BH\ and only at
$\infty$ for normal ones.\pano
So we obtain:
$$
\eqalign{
S_{BH_{non estr.}}&={{1}\over{8 \pi}} \left(\int_{\infty}
[K] -\int_{r_{h}} [K]- \int_{\infty} [K]\right) = -\int_{r_{h}} [K]
={{A}\over{4}}\cr
S_{BH_{estr.}}&={{1}\over{8 \pi}} \left(\int_{\infty}
[K] -\int_{r_{h}} [K]- \int_{\infty} [K]
+\int_{r_{h}} [K] \right) =0 \cr}
\eqn\Sbh
$$
It is well-known that one can write [\KOP]:
$$
2 \int_{\partial M} \sqrt{-h} [K ]=\int_{\partial M} \sqrt{-h}
[\omega^{\mu} n_{\mu}]
\eqn\Ko
$$
for the metrics \metr\ under our investigation one has:
$$
\eqalign{
\omega^{\mu}&=\left (0,-2e^{2U}\left(\partial_{r} U+
2 \partial_{r} \ln R \right),
-{{2 cotg \theta}\over {r^{2}}}, 0 \right) \cr
n_{\mu}&=\left( 0, {{1}\over{\sqrt{g^{11}}}},0,0 \right) \cr}
\eqn\om
$$
so:
$$
\omega^{\mu}n_{\mu}=\omega^{1}n_{1}=-2
e^{U}\left (\partial_{r} U+2\partial_{r} \ln R \right)
\eqn\omc
$$
{}From the \omc\ we have to subtract the flat metric
correspondent term:
$$
ds^{2}=-dt^{2}+dr^{2}+r^{2} d\Omega^{2}
$$
So one obtains:
$$
\eqalign{
\omega^{\mu}_{0}&=\left (0,-{4\over r},-{{2cotg \theta} \over
{r^{2}}},0 \right)\cr
n^{0}_{\mu}&=(0,1,0,0) \cr }
\eqn\os
$$
and:
$$
\eqalign{
[\omega^{\mu}n_{\mu}]&=\omega^{\mu}n_{\mu}-
\omega_{0}^{\mu}n^{0}_{\mu}=\cr
&=-2e^{U}(\partial_{r}U+2\partial_{r}\ln R)+{{4}\over{r}} \cr}
\eqn\of
$$

\chapter{Test for the espressions for $\chi$ ed $S$ for the
\sch\ space-time.
}

Here we want to check our preceding results by application to
the well-
known case of \sch\ metric:
$$
\eqalign{
e^{2U}=&(1-2M/r)\cr
U=&{1 \over 2 }\ln (1-2M/r)\cr
R=& r \cr}
$$
we find:
$$
\eqalign{
S&=-{{1}\over{8\pi}} \int_{r_{h}} d^{3}x \sqrt{-h} [K]=\cr
&=-{{1}\over{16}} \int^{\beta}_{0} d\tau e^{U} \int r^{2} \sin
\theta d \theta d \phi \left [ -2 e^{U} \left ( {{1}\over{2}}
{{2M}\over{(1-2M/r) r^{2}}}+{{2}\over{r}} \right ) +{{4}\over{r}}
\right ]_{r_{h}}\cr
&=-{{1}\over{16}} \beta 4 \pi r^{2} \left [-2 e^{2U} \left (
{{M}\over {r^{2} e^{2U}}}+{{2}\over{r}} \right )
+e^{U}{{4}\over{r}} \right ]_{r_{h}}=\cr
&={{8 \pi M}\over {4}} \left [2M+4r e^{U} \left( e^{U}-1 \right)
\right ]_{r_{h}}=\cr
&=4 \pi M^{2}={{A}\over{4}} \cr}
\eqn\en
$$
Here we used the fact that $e^{U}|_{r_{h}}=0$.\pano
So formula \of\ is exact.
We now test the formula \GBnBH\ for Euler
characteristic.\pano
For the \sch\ metric we find $\chi=2$ which is the expected
result
for this space-time of topology $R^{2} \times S^{2}$.

We can finally redo the integration of \en\ for the general case
of  a general
\sch -like metric obtaining:
$$
S=\left. {{\beta R}\over{2}} \left [
(U^{\prime}R+2R^{\prime})e^{U}-{{2R}\over{r}} \right ] e^{U}
\right|_{r=r_{h}}
\eqn\eg
$$
We can also rewrite \GBnBH\ in a more suitable form for our
next
purposes:
$$
\eqalign{
\chi&=\int_{r_{h}} \Pi={{1}\over{4 \pi^{2}}} \int_{r_{h}}
\omega^{01} \wedge R^{23}=\cr
&={{\beta}\over{2\pi}} (2 U^{\prime} e^{2U})(1-e^{2U}
{R^{\prime}}^{2} )_{r_{h}}\cr}
\eqn\chib
$$
It can be seen in this form we leave explicit the dependence on
inverse
temperature.

\chapter{ Relation between gravitational entropy and Euler
characteristic for
different no-extreme \BH\ }

{\bf { \sch\ \BH\ }}

We have:
$$
\eqalign{
e^{2U}&=(1-2M/r)\cr
U&={1\over 2 }\ln (1-2M/r)\cr
R&=r \cr}
$$
In this case $A=\beta r_{h}=4 \pi r^{2}_{h}$  and the relation
between
$\beta$ and $A$ is:
$$
\beta= \b\ ={{A}\over{r_{h}}}
$$
{}From \eg\ we obtain:
$$
S={{A}\over{4}}
\eqn\Ss
$$
We also have from \chib\ :
$$
\chi=\beta r_{h} {{1}\over {2\pi r^{2}_{h}}} ={{A}\over{2 \pi
r^{2}_{h}}}
\eqn\Cs
$$
If in \Cs\ we pose $A$ as a function of $\chi$ and substituting
it in \Ss , we
obtain:
$$
S={{\pi}\over{2} } \chi r^{2}_{h}={{\chi}\over{32
\pi}}\beta^{2}={{\chi A}\over{8}}
\eqn\SC
$$
This is the simplest relation one can imagine for the searched
relation
between entropy and Euler characteristic.
Althought it has been found for \sch\ \BH\ this different
formulation of
Bekenstein-Hawking entropy  \SC\ would give the correct
results $S=0$ for
extreme \BH\ without requiring that $A=0$.
We want to point out that this formula is valid for a large class
of \BH\
(extreme or not) and more generally for space-times described
by metric of
the type \metr\ and well defined in respect of Gauus-Bonnet
hypotesis.

{\bf  {Dilaton  $U(1)$  \BH\ with  $0 \leq a \leq 1$
(case $a=0$ correspond to Reissner-Nordstr\"om \BH\ ) } }

We have:
$$
\eqalign{
e^{2U}&=\left (1-{{r_{+}}\over{r}} \right )\left (1-{{r_{-}}
\over{r}} \right )^{{{1-a^{2}}\over{1+a^{2}}}}\cr
U&={1\over 2 }\ln \left [\left (1-{{r_{+}}\over{r}} \right )\left
(1-{{r_{-}}\over{r}} \right )^{{{1-a^{2}}\over{1+a^{2}}}}\right
]\cr
R&=r \left (1-{{r_{-}}\over{r}} \right )^{{{a^{2}}
\over{1+a^{2}}}}\cr
M&={{r_{+}}\over{2}}+{{1-a^{2}}\over{1+a^{2}}}{{r_{-}}
\over{2}}\cr
Q^{2}&={{r_{+}r_{-}}\over{1+a^{2}}} \cr
r_{h}&=r_{+}\cr}
$$
In this large class of \BH\ one finds $A=\beta R_{r_{h}}=4 \pi
R^{2}_{r_{h}}$,
$R$ determines the caracteristic scale of distance. The relation
between
$\beta$ and $A$ is:
$$
\beta= \b\ ={{A}\over
{r_{h} \left ( 1- {{r_{-}}\over{r_{h}}} \right ) }}
$$
{}From \eg\ one obtains again:
$$
S=4 \pi R^{2}_{r_{h}}={{A}\over {4}}
\eqn\Sd
$$
{}From \chib\ we also find:
$$
\chi=\beta r_{h} {{1}\over {2\pi R^{2}_{h}}} ={{A}\over{2 \pi
R^{2}_{h}}}
\eqn\Cd
$$
Putting in \Cd\ $A$ as a function of $\chi$ and inserting it
in \Sd\ we find as expected:
$$
S={{\pi}\over{2} } \chi R^{2}_{h}={{\chi A}\over{8}}
$$

{\bf {$S^{2} \times S^{2}$ (\sch -de Sitter)} }

This is another 4-dimensional gravitational instantons with a
metric of the
form \metr . This is an interesting case because there is no
boundary
$\partial M=0$ for this manifold (so \chib\ does not hold).
The Euler number is 4, topology being that of $S^{2} \times S^{2}$.
We have:
$$
\eqalign{
e^{2U}&=(1- \Lambda r^{2})\cr
U&={1\over 2 }\ln (1- \Lambda r^{2})\cr
R^{2}&=\Lambda^{-1}=cost>0 \cr
r_{h}&=\Lambda^{-1/2}\cr}
$$
By applying \SC\ to this case ($\chi=4$) one finds:
$$
S={{A \chi}\over{8}}={{4 pi \Lambda^{-1} 4}\over{8}}={{2
\pi}\over { \Lambda}}
$$
This is known to be [\EGH] the exact result.
\foot{In the work of EGH [\EGH] is given the action value
which concides
with our value for entropy (modulo a minus). This is correct
because in the
cases under our consideration the actions have the form $I=-
\alpha\beta^2$
(with $\alpha$ an opportune constant of proportionality
obtained by
esplicitating the physical dependence of system parameters
($M$ or
$\Lambda$) on temperature).
Then we have
$S=-(\beta \partial
\beta-1) I=2 \alpha \beta^2-\alpha \beta^2=\alpha \beta^2=-
I$.}
The same result can be obtained by usual path-integral
calculation.
This result implies that \SC\ holds in a more general class of
space-times
than that described by the metrics of the form \metr .

{\bf {General case}}

We have:
$$
\eqalign{
A &=4 \pi R^2 (r_h)\cr
\beta &=4\pi((e^{2U})'_{r=r_h})^{-1}\cr
S &= \left.{{\beta R}\over{2}} [(U'R+2R')e^{U}-
{{2R}\over{r}}]e^{U}\right|_{r=r_h}\cr
\chi &= \left.{{\beta}\over{2 \pi}}(2U' e^{2U})(1-
e^{2U}R'^{2})\right|_{r=r_h}\cr}
$$
so one finds:
$$
\eqalign{
S &=2 \pi \chi (2U' e^{2U})^{-1}_{r=r_h} (1-e^{2U}R'^{2})^{-
1}_{r=r_h}
{{R}\over{2}}
\left. [(U'R+2R')e^{U}-{{2R}\over{r}}]e^{U}\right|_{r=r_h}=\cr
&= \pi \chi [(e^{2U})'-R'^{2} e^{2U}(e^{2U})']^{-1}_{r=r_h} [
{{\beta
R}\over{2}}
( e^{2U})'+2R'e^{2U}-{{\beta 2R}\over{r_h}}e^{U}]_{r=r_h}\cr}
$$
by definition $\left. e^{2U}\right|_{r=r_h}=0$, so:
$$
\eqalign{
S &= \pi \chi R(r_h) \left \{ \left [ (e^{2U})' \right]^{-1} \right
\}_{r=r_h}=\cr
&= {{\pi \chi R^{2}(r_h)}\over{2}}={{\chi (4 \pi
R^{2}(r_h)}\over{8}}={{\chi
A}\over{8}} \cr}
$$

\chapter{Conclusions}

The relation \SC\ appears to hold in a wide class of space-
times.
It seems that this formulation of \BH\ entropy area possibly
clarifies the
behaviour of extreme \BH\ entropy by interpreting
gravitational entropy
has a topological effect (in this sense it confirms Hawking et al.
position
toward its interpretation.\pano
Unfortunately at the moment it appears rather difficult to find
a dynamical
explanation of this ``topological" entropy. We conjecture that
the shown deep
relation  of this entropy with boundary structure of the space-
time is in a
certain sense a hint towards an interpretation based on
dynamical degrees of
freedom associated to vacuum in topological non trivial space-
times
\Ref\mes{S.~Hacyan, A.~Sarmiento, G.~Cocho, F.~Soto\journal Phys. Rev. D&32
(1985) 914.}.\pano
Maybe that intrinsic thermodynamics of some
gravitational instantons is statistically due to Casimir-like
effects of vacuum.
Zero-modes are known to be sensible to the topological
structure of
space-time but it is also not well understood how to associate
them a
termodinamical interpretation.

{\bf {Note added}}

When this work in its main results has been already completed, the
authors became
aware of recent work by
\Ref\BTZ{Banados, Teitelboim, Zanelli\journal
Phys.Rev.Lett&72, 7 (1994)
957.} and
\Ref\BTZ{C.Teitelboim - {\em Action and entropy of extreme
and non-extreme
black holes} - Preprint hep-th 9410103.} in which the relation
between entropy and
topology is put in evidence. In our opinion this work does not
have a
substantial overlapping with these already quoted.
These are based on hamiltonian formulation of the problem
and the
thermodynamical quantities are espressed as function of the
Euler number of
a manifold which is not the physical one (but of course the
respective results
are in complete accord).

\ACK{S.Liberati wish to thank F. Belgiorno for some
illuminating remarks; G. Esposito, G. Immirzi, M. Maggiore,
M. Martellini, K. Yoshida  for some constructive advices.
The authors are also gratefull to P. Blaga,
B. Jensen and R.Garattini for extensive discussions.}
\endpage\refout
\end